\input epsf
\documentstyle{mn}
\newif\ifAMStwofonts
\def\lesssim{\mathrel{\hbox{\rlap{\hbox{\lower4pt\hbox{$\sim$}}}\hbox{$<$}}}}
\def\gtrsim{\mathrel{\hbox{\rlap{\hbox{\lower4pt\hbox{$\sim$}}}\hbox{$>$}}}}
\def\apj{ApJ}

\def\aap{A\&\hskip-1pt A}

\def\mnras{MNRAS}
\def\pasp{PASP}

\title[Analytic Relations of Blending]
      {Analytic Relations between the Observed Gravitational \\
      	Microlensing Parameters With and Without the Effect of Blending}
\author[Cheongho Han]
       {Cheongho Han\\
        Dept.\ of Astronomy \& Space Science, \\
        Chungbuk National University, Chongju, Korea 361-763\\
	cheongho@astronomy.chungbuk.ac.kr}

\date{Accepted
      Received }
\pagerange{\pageref{firstpage}--\pageref{lastpage}}
\begin{document}

\maketitle

\label{firstpage}

\begin{abstract}
When a microlensing light curve is contaminated by blended light from 
unresolved stars near the line of sight to the lensed star, the light 
curve shape and corresponding parameterization for the event will differ 
from the values expected when the event is not affected by blending.  
As a result, blending makes it difficult to identify the major lens 
population and to estimate the amount of lensing matter.  In order to 
estimate the effect of blending on the result of lensing experiments, it is, 
therefore, essential to know how the observed lensing parameters change 
depending on the fraction of blended light.  Previously, the changed 
lensing parameters were obtained with a statistical method that not only 
required a large amount of computation time but also was prone to uncertainty.
In this paper, we derive analytic relations between the lensing parameters 
with and without the effect of blending.  By using these relations, we 
investigate the dependence of the observed lensing parameters on the amount 
of blended light, the impact parameter, and the threshold amplification for 
event detection.
\end{abstract}


\section{Introduction}
Following the proposal of Paczy\'nski (1986), searches for Massive 
Astronomical Compact Objects (MACHOs) by detecting light variations of 
stars caused by gravitational microlensing are being carried out by several 
groups (MACHO, Alcock et al.\ 1997; EROS, Ansari et al.\ 1996; OGLE, Udalski
et al.\ 1997).

When an isolated source star is gravitationally microlensed, 
its flux is amplified by an amount
$$
A_0(u) = {u^2+2 \over u\left( u^2+4\right)^{1/2}};\qquad
u = \left[ \beta^2 + \left( {t-t_0\over t_{\rm E}}\right)^2\right]^{1/2},
\eqno(1.1)
$$
where $u$ is the lens-source separation in units of the angular 
Einstein ring radius, $\theta_{\rm E}$.  The lensing parameters 
$t_0$, $\beta$, and $t_{\rm E}$ represent the time of maximum 
amplification, the lens-source impact parameter in units of 
$\theta_{\rm E}$, and the Einstein ring radius crossing time 
(Einstein time scale), respectively. These parameters are obtained 
by fitting model light curves in equation (1.1) to the observed one.  
Among these parameters, the Einstein time scale is related to the 
physical parameters of the lens by
$$
t_{\rm E} = {r_{\rm E}\over v};\qquad
r_{\rm E} = \left({4GM \over c^2} {D_{ol}D_{ls}\over D_{os}}\right)^{1/2},
\eqno(1.2)
$$
where $M$ is the mass of the lens, $v$ is the lens-source transverse
velocity, $D_{ol}$, $D_{ls}$, and $D_{os}$ are the separations between 
the observer, lens, and source star, and $r_{\rm E}=D_{ol}\theta_{\rm E}$ 
is the physical size of the Einstein ring radius.  Since the Einstein 
time scale results from the combination of the physical parameters of the 
lens, it is difficult to obtain information about these parameters of 
individual lenses.  However, one can still obtain information about 
the major population of lenses from the distribution of Einstein time 
scales by modeling the distribution and motion of lens matter constrained 
from other types of observation (Han \& Gould 1996; Gould 1996; Zhao, Rich, 
\& Spergel 1996; Han \& Chang 1998).  In addition, from the comparison of 
the experimentally determined optical depth, which is proportional to the 
summation of the Einstein time scales of individual events, i.e.\ 
$\tau\propto\sum_i t_{{\rm E},i}$, to the theoretically expected value, 
experiments toward Magellanic Clouds allow one to constrain the MACHO dark 
matter fraction in the Galactic halo.

However, the observed light curves of a significant fraction of lensing 
events will differ from the light curve in equation (1.1).  This is because 
current lensing experiments are being conducted toward very dense star 
fields such as the Magellanic Clouds and the Galactic bulge.  While searches 
towards these crowded fields result in an increased event rate, it also 
implies that many of the observed light curves are blended by the light from 
unresolved stars that are not being lensed.  When an event is affected by 
blending, the observed light curve becomes 
$$
A_{\rm obs}(u,B/F_0) = {A_0(u)F_0+B\over F_0+B} 
                     = {A_0(u)+B/F_0\over 1+B/F_0},
\eqno(1.3)
$$
where $F_0$ is the unblended and unamplified flux of the lensed star and 
$B$ is the amount of blended light.  The problem of blending is that it is 
very difficult to detect the presence of a blend, and practically impossible 
to correct for the effect by purely photometrical means (Wo\'zniak \& 
Paczy\'nski [1997], see also Figure 2 for example blended light curves and 
the corresponding degenerate unblended light curves).  Since blending makes 
the observed parameters of an event differ from the values expected when the 
event is not affected by blending, blending  leads to misidentification of 
the major lens population.  In addition, the MACHO fraction determined from 
the lensing optical depth will be subject to great uncertainty.  Throughout 
this paper, we consistently use term `observed lensing parameters' to 
represent the lensing parameters determined from a blended light curve not 
knowing it is affected by blending.

Blending can be classified into several types depending on the origin 
of the blended light.  The first type, regular blending, occurs when 
a bright source star registered on a template plate is lensed and its 
flux is affected by the light from faint unresolved stars below the 
detection limit imposed by crowding.  Second, lens blending occurs when 
the origin of the blended light is the lens itself (Kamionkowski 1995; 
Buchalter \& Kamionkowski 1997; Nemiroff 1997; Han 1998).  Third, 
amplification-bias blending occurs when one of several unresolved faint 
stars in the seeing disk of a reference star is lensed, and the flux 
of the lensed star is associated with the flux from other stars in 
the integrated seeing disk (Nemiroff 1994; Han 1997a; Alard 1997).  Finally, 
if a source is composed of binaries and only one of the component is 
gravitationally amplified, the light from the other binary component simply 
contribute as blended light: binary-source blending (Griest \& Hu 1992; 
Dominik 1998; Han \& Jeong 1998).

In order to estimate the effects of the individual types of blending on 
the results of lensing experiments, one has to know how the observed 
lensing parameters change depending on the fractions of blended light.
One of the methods to determine the changed lensing parameters  
is statistically fitting a series of unblended light curves to the
blended light curve of interest (e.g., Wo\'zniak \& Paczy\'nski 1997).
In this case, determining the parameters requires a large amount of 
computation time for fitting process.  Furthermore, for the full analysis 
of the blending effect, one should repeat the same procedure for events 
with all combinations of lensing parameters.  In addition, there are 
inevitable statistical uncertainties that accompany the use of this method. 
Di Stefano \& Esin (1995) attempted to develop analytic formalism for the 
treatment of blending.  However, their so-called ``blended Einstein ring 
radius'' quantifies only how the maximum allowed impact parameter of 
lens-source encounter for event detection changes as a function of blended 
light fraction (see \S\ 2).  Since all information about lenses are obtained 
from lensing parameters, especially the Einstein time scale, simple analytic 
relations between the lensing parameters with and without blending will be 
very useful for the estimation of blending effect on the result of lensing 
experiments.

In this paper, we derive these relationships and use the relations to 
investigate the dependence of the observed lensing parameters on the 
amount of blended light, the impact parameter, and the threshold 
amplification for event detection.

\section{Changes in Lensing Parameters Due to Blending}

The shape of a microlensing event light curve is characterized by its 
height (peak amplification $A_p$) and width (event duration $t_d$), which 
are dependent on the lensing parameters: $\beta$ for the height (see 
equation [2.2]) and $t_{\rm E}$ and $\beta$ for the width (see equation 
[2.5]).  Because of the change in the peak amplification due to blending, 
the observed impact parameter differs from the value determined from 
unblended light curve.  Without blending, the impact parameter is related 
to the peak amplification of the unblended light curve by 
$$
\beta = 
\left[ 2\left( 1-A_{p}^{-2}\right)^{-1/2}-2\right]^{1/2}.
\eqno(2.1)
$$
When the event is affected by blending, on the other hand, 
the observed peak amplification decreases into
$$
A_{p,obs}(\beta,B/F_0) = { A_{p}(\beta)+B/F_0\over 1+B/F_0};
\qquad
A_{p}(\beta) = {\beta^2 + 2 \over \beta(\beta^2+4)^{1/2}}.
\eqno(2.2)
$$
As a result, the impact parameter obtained by ignoring blending effect
becomes dependent on the amount of the blended light by
$$
\beta_{obs}(\beta,B/F_0) = 
\left[ 2\left( 1-A_{p,obs}^{-2}\right)^{-1/2}-2\right]^{1/2}.
\eqno(2.3)
$$

%Figure 1
\begin{figure}
%\vspace{302pt}
\epsfysize=10cm
\centerline{\epsfbox{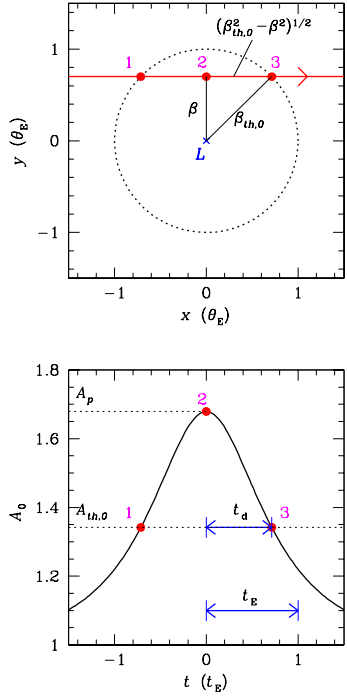}}
\caption{
Upper panel: The lens system geometry of an unblended lensing event.  
To be detected a source star should enter the ring (dotted line) around 
the lens ({\it L}, marked by `x') with a radius $\beta_{th,0}$ (threshold 
impact parameter).  If the adopted threshold amplification is 
$A_{th,0}=3/\sqrt{5}$, $\beta_{th,0}=1$, and the ring is equivalent 
to the Einstein ring.  The straight line (with an arrow) represents 
a source star trajectory.  The three points on the source star 
trajectory (marked by numbers) represent the source star positions at 
the moments when it enters the ring, approaches most closely to the 
lens, and leaves the ring, respectively.  Lower panel: The resulting 
light curve from the lens system geometry in the upper panel.  
The three points on the light curve (marked also by numbers) 
represent the amplifications at the source star positions with 
corresponding numbers in the upper panel.  The peak amplification 
($A_p$) and the threshold amplification ($A_{th,0}$) are represented 
by dotted lines.  The Einstein time scale ($t_{\rm E}$) and the event 
duration ($t_{d}$) are marked by thick solid lines with arrows.
}
\end{figure}

Not only the peak amplification, but also the event duration is affected 
by blending.  As an event is always amplified by more than 1 by definition, 
we define the event duration by half of the time a given event has an 
amplification greater than some threshold amplification of $A_{th,0}$. 
\footnote{
To avoid confusion, we note the difference between the event duration,
$t_{d}$, and the Einstein time scale, $t_{\rm E}$.  For an unblended 
event detected with a 
threshold amplification of $A_{th,0}=3/\sqrt{5}$, $t_{d}$ represent 
the half of time for the source to stay within the Einstein 
ring while $t_{\rm E}$ is the time scale for a source to cross the 
Einstein ring radius.  The longest source star trajectory within the 
Einstein ring is the Einstein ring diameter.  Therefore, the two time 
scales become equal only when the source crosses the center of the 
Einstein ring, but otherwise $t_{d}$ is always less than $t_{\rm E}$.
}
For a given threshold amplification for event detection,
the maximum allowed impact parameter (threshold impact parameter) is 
determined by
$$
\beta_{th,0}(A_{th,0}) = 
\left[ 2\left( 1-A_{th,0}^{-2}\right)^{-1/2}-2\right]^{1/2}.
\eqno(2.4)
$$
For example, if the adopted threshold amplification is $A_{th,0} = 
3/\sqrt{5}$, the threshold impact parameter becomes $\beta_{th,0}=1$.
According to this definition of a detectable microlensing 
event, the source star should enter the Einstein ring.  However, since 
the threshold amplification can be different values depending on the 
observational strategy and corresponding event selection criteria, we 
leave the threshold amplification as a variable.  From the definition 
of the threshold impact parameter as 
$\beta_{th,0}^2 = \beta^2 + \left( t_{d}/ t_{\rm E}\right)^2$, 
one finds that the duration of an unblended event is related to the 
threshold amplification and the unblended lensing parameters by
$$
t_{d}(A_{th,0},t_{\rm E},\beta) 
= t_{\rm E}\left( \beta_{th,0}^2-\beta^2\right)^{1/2}.
\eqno(2.5)
$$
In Figure 1, to visualize the definitions of $\beta_{th,0}$, $A_{p}$,
and $t_{d}$ and the relation between $t_{\rm E}$ and $t_{d}$, we  
present the geometry of a lens system for an unblended event (upper panel)
and the resulting light curve (lower panel).  In the upper panel, the source 
star trajectory is represented by a straight line.  In the lower panel, 
the peak amplification ($A_{p}$) and the threshold amplification ($A_{th,0}$) 
are represented by dotted lines.  The Einstein time scales ($t_{\rm E}$) 
and the event duration ($t_{d}$) are marked by thick solid lines with arrows.

On the other hand, if the event is affected by blending, the same event 
must have a higher amplification because blending increases the threshold 
amplification to: 
$$
A_{th}(A_{th,0},B/F_0) = A_{th,0}(1+B/F_0) - B/F_0.
\eqno(2.6)
$$
Blending also decreases the observed duration of the event to:
$$
t_{d,obs}(A_{th,0},t_{\rm E},\beta,B/F_0) 
= t_{\rm E}\left( \beta_{th}^2-\beta^2\right)^{1/2},
\eqno(2.7)
$$
where 
$$
\beta_{th} (A_{th,0},\beta,B/F_0)= 
\left[ 2\left( 1-A_{th}^{-2}\right)^{-1/2}-2\right]^{1/2}
\eqno(2.8)
$$
is the decreased threshold impact parameter caused by the increased 
threshold amplification $A_{th}$ in the presence of blending. 
We note that the formula of so-called ``blended Einstein ring'' of 
Di Stefano \& Esin (1995) is identical to our threshold impact parameter 
in equation (2.8) for a special case for $A_{th,0}=3/\sqrt{5}$.

As a result of the changes in the impact parameter and the duration 
of event, the Einstein time scale also changes.  Not knowing an event is 
affected by blending, one will obtain the Einstein time scale from the 
relation in equation (2.5), but based on the observed impact parameter and 
the duration of event by 
$$
t_{{\rm E},obs}= { t_{d,obs}\over 
\left( \beta_{th,0}^2-\beta_{obs}^2 \right)^{1/2}}.
\eqno(2.9)
$$
In the equation, the threshold impact parameter $\beta_{th,0}$ is included 
instead of the decreased value of $\beta_{th}$ due to blending, because one 
still thinks he (or she) consistently applied the same threshold impact 
parameter of $\beta_{th,0}$.  Since the observed duration of the event is 
related to the unblended lensing parameters by equation (2.7), one obtains the 
relation between the Einstein time scales with and without blending by
$$
t_{{\rm E},obs} = t_{\rm E} \left( \
{\beta_{th}^2 - \beta^2\over
\beta_{th,0}^2 - \beta_{obs}^2}
\right)^{1/2},
\eqno(2.10)
$$
where the definitions of the various types of impact parameter of $\beta$, 
$\beta_{obs}$, $\beta_{th,0}$, and $\beta_{th}$ are given in equations 
(2.1), (2.3), (2.4), and (2.8).

%Figure 2
\begin{figure}
%\vspace{302pt}
\epsfysize=9.5cm
\centerline{\epsfbox{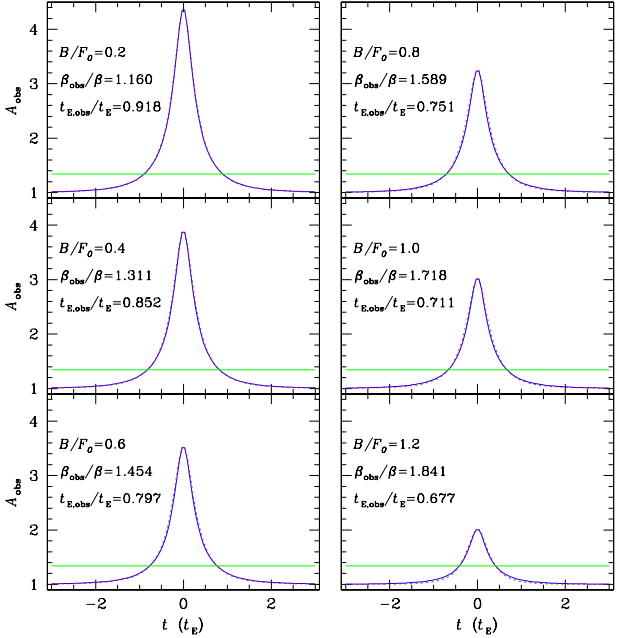}}
\caption{
Blended event light curves (solid curves) and the corresponding 
degenerate unblended light curves (dotted curves).  The lensing 
parameters determined from the unblended fit to the blended light 
curves (i.e., $\beta_{obs}$ and $t_{{\rm E},obs}$) are obtained by using 
the derived equations (2.3) and (2.10).  All of these events have the 
same unblended impact parameter ($\beta=0.2$) and the Einstein time scale.  
The adopted threshold amplification is $A_{th,0}=3/\sqrt{5}$ (represented 
by a solid line).  The fraction of blended light ($B/F_0$) and the fractional 
changes in the lensing parameters ($\beta_{obs}/\beta$ and 
$t_{{\rm E},obs}/t_{\rm E}$ for the impact parameter and the Einstein time 
scale, respectively) with and without blending are marked 
in each panel.  One finds that the blended light curves are very well 
fit by unblended light curves of different lensing parameters.
}
\end{figure}

In Figure 2, we present several illustrative light curves with blending 
(solid curves) and the corresponding degenerate unblended light curves 
(dotted curves) whose lensing parameters (i.e., $\beta_{obs}$ and 
$t_{{\rm E},obs}$) are determined by our derived relations in equations (2.3) 
and (2.10).  All of these events have the same unblended impact parameter 
$\beta=0.2$ and the adopted threshold amplification is $A_{th,0}=3/\sqrt{5}$ 
(represented by a solid line).  The fraction of blended light and the 
fractional changes in the lensing parameters with and without blending are 
marked in each panel.  As shown by Wo\'zniak \& Paczy\'nski (1997), the 
light curves of blended events are very well fit by unblended light curves 
of different lensing parameters.

With the derived analytic relations, the effects of blending on the observed
lensing parameters can be computed explicitly.  We compute these changes 
in the lensing parameters due to blending with the equations derived above 
for the impact parameter (2.6) and for the Einstein time scale 
(2.10). Figure 3 shows how these parameters vary as a function of the 
blended light fraction for different values of the impact parameter. 
To compute the observed lensing parameters, we assume that the threshold 
amplification for event detection is $A_{th,0}=3/\sqrt{5}$.  In the figure, 
the relation becomes discontinuous above a certain value of $B/F_0$ 
because events affected by more blending than this amount cannot be 
amplified higher than $A_{th,0}$ and thus cannot be detected.  As expected, 
one finds that the observed Einstein time scale decreases and the observed 
impact parameter increases as the fraction of blended light increases. 
We note, however, that although these qualitative trends of the changes 
in the observed lensing parameters due to blending are already known 
(Di Stefano \& Esin 1995; Alard \& Guibert 1997; Han 1997b; Goldberg 
1998) and numerical quantification based on statistical methods was 
performed (Wo\'zniak \& Paczynski 1997), our derivation is the first to 
analytically quantify these changes.

Another interesting result from Figure 3 is that the observed lensing 
parameters depend in different ways on $B/F_0$ for different values 
of the unblended impact parameter $\beta$.  This is because even for events
with the same Einstein time scale and the same fraction of blended light, 
the peak amplification and the duration of the event increase as the impact 
parameter decreases. The decrease in $t_{{\rm E},obs}$ and the increase in 
$\beta_{obs}$ become more important for events with smaller impact 
parameters, implying that blending is more important for events with 
higher amplifications.

%Figure 3
\begin{figure}
%\vspace{302pt}
\epsfysize=9.5cm
\centerline{\epsfbox{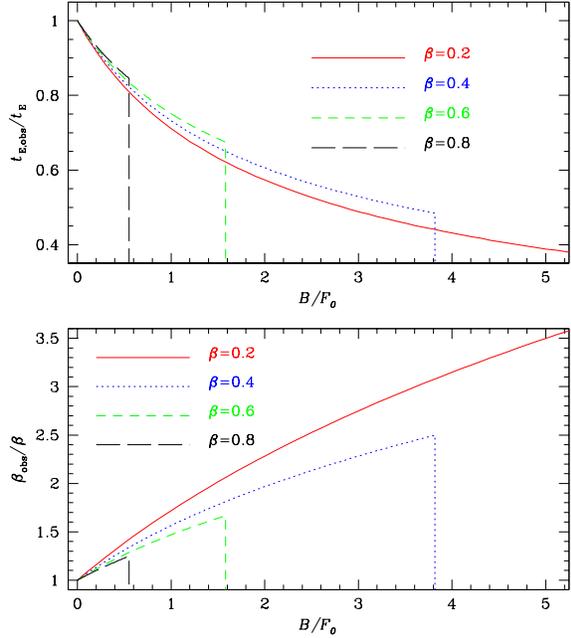}}
\caption{
The fractional changes in the observed lensing parameters 
($t_{{\rm E},obs}/t_{\rm E}$ and $\beta_{obs}/\beta$ for the Einstein time
scale and the impact parameter, respectively) as a function of the 
fraction of blended light ($B/F_0$) for events with different values of the 
impact parameter.  We assume that the threshold amplification for event 
detection is  $A_{th,0} = 3/\sqrt{5}$.  In the figure, the relation 
becomes discontinuous at a certain value of the blended light fraction 
$B/F_0$ because events more strongly affected by this amount of blended 
light cannot be amplified higher than 
$A_{th,0}$, and would fall below the detection threshold.
}
\end{figure}

%Figure 4
\begin{figure}
%\vspace{302pt}
\epsfysize=9.5cm
\centerline{\epsfbox{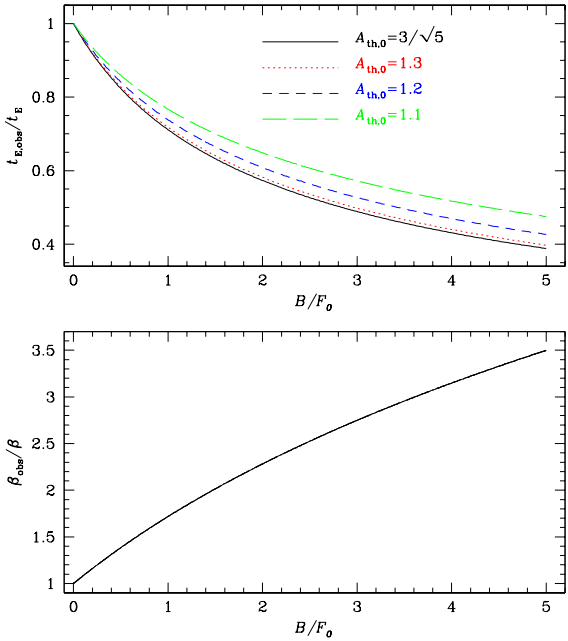}}
\caption{
The dependence of the observed lensing parameters 
($t_{{\rm E},obs}/t_{\rm E}$ and $\beta_{obs}/\beta$ for the Einstein time
scale and the impact parameter, respectively) on the fraction of blended 
light ($B/F_0$)  for different values of the adopted threshold amplification 
$A_{th,0}$.  This model event has an unblended impact parameter of 
$\beta=0.2$. As the value of the threshold amplification increases, the 
decrease in the observed Einstein time scale becomes more important.  
Unlike the Einstein time scale, however, the observed impact parameter 
has the same dependence on $B/F_0$ regardless of the adopted values of 
$A_{th,0}$.  Note that the four lines in the lower panel coincide.
}
\end{figure}

We also find that the observed Einstein time scale depends differently on 
the fraction of blended light for different values of the adopted threshold 
amplification.  Figure 4 shows how $t_{{\rm E},obs}$ changes for different 
values of $A_{th,0}$ for an example event with $\beta=0.2$.  From the figure, 
one finds that as the threshold amplification increases, the decrease in the 
observed Einstein time scale becomes more important.  This is because by 
increasing the threshold amplification the observed duration of an event 
decreases, while the unblended Einstein time scale remains the same regardless 
of the adopted $A_{th,0}$.  Unlike the Einstein time scale, however, the 
observed impact parameter has the same dependency on $B/F_0$ regardless of the 
adopted values of $A_{th,0}$.  This is because $\beta_{obs}$ is determined 
solely from the peak amplification, which does not depends on the adopted 
$A_{th,0}$.

\section{Conclusion}

We have derived analytic relations for the observed microlensing parameters    
with and without the consideration of blending effect. We have used these 
relations to investigate the dependence of the observed lensing parameters on 
the fraction of blended light for events with various values of impact 
parameters under various selection criteria.  
The results of this investigation are as follows: 

\begin{enumerate}
\item
The observed Einstein time scale decreases and the impact parameter increases 
with an increasing fraction of blended light. Although these trends in the 
observed lensing parameters are already known, this derivation is the first 
to analytically quantify this effect. 

\item 
The observed lensing parameters depend in different ways on the
fraction of blended light for events with different impact parameters.
These changes in the observed lensing parameters are more important for
events with smaller impact parameters, implying that the effects of blending
becomes more important for events with higher amplifications.
\item 
The observed Einstein time scale depends on the threshold amplification for 
event detection.  With increasing value of the adopted threshold amplification, 
the decrease in the observed Einstein time scale becomes more important.  
However, the observed impact parameter has the same dependence regardless of 
the adopted threshold amplification.
\end{enumerate}

\section*{Acknowledgments}
We would like to thank P.\ Martini for a careful reading of the manuscript.

\end{document}